\documentclass[useAMS, usenatbib]{mn2e}
\usepackage{graphicx,amssymb,amsmath}
\usepackage{hyperref,xcolor}
\usepackage{subfigure}
\usepackage{pdflscape}
\usepackage{mathrsfs}

\usepackage[T1]{fontenc}
\begin{document}






\title[Mutual events 2014-15 analysis]{Analysis of mutual events of Galilean satellites observed from VBO during 2014-15}

\author[R. Vasundhara et al.]{
R. Vasundhara,$^{1}$\thanks{E-mail:rvas@iiap.res.in(RV),selva@iiap.res.in(GS), anbu@iiap.res.in(PA)}
G. Selvakumar,$^{2}$
and P. Anbazhagan$^{2}$
\\
$^{1}$Indian Institute of Astrophysics, Bangalore 560034, India\\
$^{2}$Vainu Bappu Observatory, VBO, Indian Institute of Astrophysics, Kavalur, India\\
}

\date{Accepted XXX. Received YYY; in original form ZZZ}

\pubyear{2016}

\label{firstpage}
\pagerange{\pageref{firstpage}--\pageref{lastpage}}
\maketitle

\begin{abstract}
Results of analysis of 23 events of the 2014-2015 mutual event series from the 
Vainu Bappu Observatory are presented. Our intensity distribution model for
the eclipsed/occulted satellite is based on the criterion that it simulates a 
rotational light curve that matches the ground based light curve. Dichotomy 
in the scattering characteristics of the leading and trailing sides explain 
the basic shape of the rotational light curves of Europa, Ganymede and Callisto. 
In case of Io the albedo map from USGS along with global values of scattering 
parameters work well. Mean values of residuals in $(O-C)$ along and perpendicular 
to the track are found to be -3.3 and -3.4 mas respectively compared to "L2" 
theory for the seven 2E1/2O1 events. The corresponding R.M.S values are 8.7 and 
7.8 mas respectively. For the five 1E3/1O3 events, the along and perpendicular to 
the track mean residuals are 5.6 and 3.2 mas respectively. The corresponding 
R.M.S. residuals are 6.8 and 10.5 mas respectively. We compare the results using 
the chosen model (Model 1) with a uniform but limb darkened disk (Model 2). The 
residuals with Model 2 of the 2E1/2O1 and 1E3/1O3 events indicate a bias along the 
satellite track. The extent and direction of bias is consistent with the shift of 
the light center from the geometric center. Results using Model 1, which 
intrinsically takes into account the intensity distribution show no such bias.
\end{abstract}




\section{Introduction}
On the basis of the results of 
the international campaign  in 1973 of the mutual events 
of the Galilean satellites, \citet{Aksnes1976} 
showed that differences in the observed and
predicted mid times and depths of the events can be used to revise the orbital constants 
of the theory. Subsequent International campaigns  and analysis of the  observations during 
the 1979, 1985, 1991, 1997, 2003 and 2009 apparitions have provided an
extensive data set 
of inter satellite astrometric positions 
\citep{Aksnes1984,GSO1991,Vasundhara1991,Arlot1992,Mallama1992,Vasundhara1994,Arlot1997,
kaas1999,Vasundhara2002,Vasundhara2003,Arlot2006,Arlot2009,Emelyanov2009,Dias-Oliveira2013,Arlot2014}. 
Following the success of the jovian mutual events, 
the technique has been extended to mutual events of saturnian
system in 1995 and 2009 \citep{Thuillot2001,Mallama2009, Arlot2012, Zhang2013}, and uranian 
system in 2007  
\citep{Assafin2009,Christou2009,Arlot2013}. We report here astrometric results of 23 mutual 
events of the jovian satellites observed from the Vainu Bappu Observatory(VBO) during 2014-15.

\section{Observations and data reduction}
    The observations were carried out at the 1.3 m J.C.Bhattacharyya telescope 
which is a F/8 R-C Double Horseshoe telescope by DFM Engineering. The 
ProEM CCD detector covers a field of $4\arcmin \times 4 \arcmin$ at 0.26 arcsec/pixel resolution.
A field of this size enabled registering a comparison satellite in the CCD frame along with the
main objects on most occasions. 
The frames were read at 1 MHz speed in normal mode. The time was read from Telescope 
control system (GPS time derived through Telescope control software by DFM) at the start 
of data acquisition ($UT_{Acqn}$) and was put in the FITS header. The exposure start 
and end instances available from the Lightfield program are accurate up to 
1 microsecond. However, due to the delay in 
accessing  $UT_{Aqun}$ from the telescope control system by few tens of microseconds, 
the UT times are only accurate to 100 $\mu s$. The final time 
scale was converted to UTC at the time of data reduction. 
Sky conditions were generally poor and only those observations that had a comparison 
satellite in the CCD frame were found to be useful. 
Images were obtained  before and/or after the event when the satellites
were well separated in case of occultation events and eclipse events close to
opposition.The derived ratio $r$ given by 
\begin{equation}
   r = \frac{IS1}{IS1 + IS2}  
        \label{eq:Eq1}
\end{equation}
was used to account for the contribution of the occulting or eclipsing satellite if
present in the aperture. In Equation \ref{eq:Eq1}, IS1 and IS2 are the intensities of 
the eclipsing/occulting
and eclipsed/occulted satellites respectively outside the event. 
The observed normalized light curve with
the contributions from  both S1 and S2 is:
\begin{equation}
F(S1,S2,t)=\frac{I(t)}{I(0)},
  \label{eq:S1S2Flux}
\end{equation}
where I(t) is the combined intensity of S1 and S2 at time "t" during the event and $I(0)=IS1+IS2$.
The observed normalized light curve
after removing the contribution of S1 can be written as:
\begin{equation}
f(S2,t)=\frac{I(t)-I(0)r}{I(0)-I(0)r}.
  \label{eq:S2Flux}
\end{equation}
Our model computes the intensity variation $f^{C}(S2,t)$ of S2 during the event. 
This model light curve could in principle be fitted with $f(S2,t)$ directly.
However, we preferred to keep the observed light curve intact and
converted instead the computed normalized intensity $f^{C}(S2,t)$ 
to $F^{C}(S1,S2,t)$ using the following relation:
\begin{equation}
F^{C}(S1,S2,t)=f^{C}(S2,t)(1-r)+r.
  \label{eq:S1S2Fit}
\end{equation}
The observed light curve $F(S1,S2,t)$ and the computed light curve
$F^{C}(S1,S2,t)$ were fitted as shown in the next section.
Standard {\it R} filter was used for all the events. 
The flux of the  objects were extracted using aperture photometry in the
APPHOT/DIGIPHOT package in IRAF. 
Table \ref{tab:Obs_table} gives
the observing conditions of the twenty three  events that were found to be 
usable for analysis. The integration time was
selected between 0.2 - 2 seconds depending on the sky conditions. Column 6
gives the time interval between the data points. The 
value of 'Q' in column seven is '0' for complete light curves 
and '1' for light curves for which either immersion  or emersion is illdefined. 
The "ratio" computed using Equation~\ref{eq:Eq1}
is given in column 8. A value of zero indicates presence
of only the eclipsed/occulted satellite in the aperture during flux extraction.

\begin{table*}
	\centering
	\caption{Details of Observations}  
	\label{tab:Obs_table}
	\begin{tabular}{cccccccc} 
		\hline
    UT Date  & Event&Comp. &Seeing&Airmass&Cycle time&Q& $Ratio, r$\\
  yyyy mm dd &      &      &arcsec&       &sec       & \\
(1)&(2)&(3)&(4)&(5)&(6)&(7)&(8)\\
		\hline
  2014 11 27 & 2O4 &J3 &2.2 &1.142&1.6405&0& 0.6624$\pm$0.0040\\
  2015 01 26 & 3E1 &J3 &2.5 &2.333&1.6405&0& 0.0000\\
  2015 01 29 & 1E3 &J1 &2.2 &1.586&1.5405&0& 0.0000\\
  2015 01 31 & 2E1 &J4 &2.1 &1.002&1.3407&0& 0.4194$\pm$0.0136\\
  2015 01 31 & 2O1 &J4 &2.0 &1.002&1.3407&0& 0.4194$\pm$0.0136\\
  2015 01 31 & 2E4 &J1 &2.1 &1.075&1.4406&0& 0.0000\\
  2015 02 12 & 1E3 &J1 &3.1 &1.309&1.3102&1& 0.0000\\
  2015 02 18 & 2E1 &J3 &5.0 &2.641&3.1408&0& 0.4290$\pm$0.0059\\
  2015 02 25 & 2O1 &J3 &2.4 &1.260&1.3407&0& 0.4319$\pm$0.0024\\ 
  2015 02 25 & 2E1 &J3 &2.1 &1.173&1.6405&0& 0.0000\\
  2015 02 26 & 4O2 &J3 &1.9 &1.304&1.4406&0& 0.2748$\pm$0.0046\\
  2015 03 07 & 2E4 &J3 &1.9 &1.173&1.6405&0& 0.0000\\
  2015 03 11 & 2O1 &J3 &1.9 &1.259&1.3407&0& 0.4396$\pm$0.0047\\
  2015 03 11 & 2E1 &J3 &2.0 &1.725&1.4406&0& 0.0000\\
  2015 03 21 & 2E3 &J2 &1.8 &1.008&4.0220&0& 0.0000\\
  2015 03 24 & 2E4 &J2 &2.5 &1.330&1.7000&0& 0.0000\\
  2015 03 28 & 1E3 &J1 &1.8 &1.015&1.3102&0& 0.0000\\
  2015 03 28 & 2E3 &J2 &2.2 &1.569&1.6100&0& 0.0000\\
  2015 04 02 & 4E3 &J4 &2.2 &1.779&1.6100&0& 0.0000\\
  2015 04 04 & 1O3 &J2 &1.6 &1.005&1.4406&0& 0.4211$\pm$0.0017\\
  2015 04 05 & 2E1 &J2 &2.0 &1.079&4.6216&0& 0.0000\\
  2015 04 11 & 1O3 &J2 &2.4 &1.500&1.6100&1& 0.4219$\pm$0.0029\\
  2015 06 03 & 3O1 &J2 &1.8 &1.900&1.6405&1& 0.5686$^1$\\
		\hline
		\hline
\multicolumn{8}{l}{1. From published light curve \citep{MorrisonMorrison} in V magnitude.}
	\end{tabular}
\end{table*}

\section{Model fit to the observations} 
\label{sec:Model}
\subsection{Ephemerides}
\label{subsec:Ephem}
Ephemerides of Jupiter, the Earth and the Sun were
computed using SPICE kernels \citep{Acton1996} based on DE430. 
The positions of the Galilean satellites with respect to Jupiter barycenter were calculated using 
the theory "L2"  by \citet{Lainey2009}; the required kernel 
NOE-5-2010-GAL-a2.bsp available at SPICE was used.
\subsection{Selection of albedo and limb darkening model for the satellites}
\label{subsec:Albedo}
In order to
exploit the full potential of the mutual events which are capable of yielding high positional accuracy, 
it is important to use a realistic intensity
distribution on the satellites. 
  The Galilean satellites are known to have  rotational light curves
implying non uniformity in albedo over their surface \citep{Stebbins1927,Stebbins1928,
MorrisonMorrison}.
\citet{Aksnes1976} pointed out the importance of including surface
variations because some light curves were asymmetric giving clear  indication
that such brightness variations do exist. 
\citet{Vasundhara2002} 
used the mosaics of the Galilean satellites constructed by the USGS team 
and estimated that for central events, the shifts of the photo-center were
in the range $-50$ to $+90$ km ($\approx -14$ to $+25$ mas) for Io and
$-30$ to $+50$ km ($\approx$ $-8$ to $+14$ mas) for Europa. 
\citet{EmelyanovGilbert2006} 
used the albedo maps of the satellites to calculate integrated brightness of
the satellites at different rotational phases. They
point out that while there are insignificant changes between their modeled 
and observed rotational light curves  by \citet{MorrisonMorrison}
 for Io and Europa, there is deviation for Ganymede.
In case of Callisto they find significant discordance. \citet{Prokofjeva2012}
point out that the spaceborne observations were carried out at a wide range of phase angles
while the groundbased observations are limited to 0$^{\circ}$ - 12$^{\circ}$. They
attribute the reason for the discrepancy between groundbased and spaceborne rotational light curves
of Ganymede and Callisto to back scattering. 
It is therefore essential 
to chose a realistic intensity distribution model. We follow the suggestion by 
\citet{EmelyanovGilbert2006} that the selected distribution
should follow the groundbased rotational
light curve of the satellites. 

\begin{figure}
\hskip -0.3cm
        \includegraphics[width=3.5in]{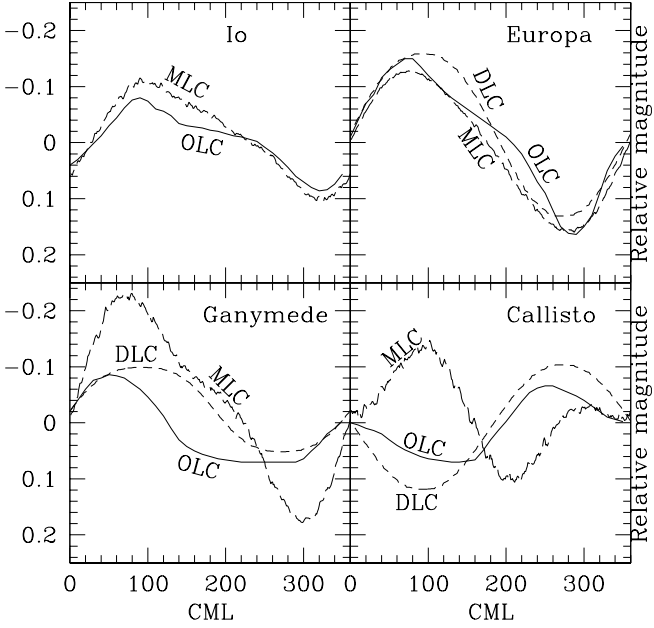}
    \caption{Comparison of rotational light curves. 
OLC: V magnitude light curve by \citet{MorrisonMorrison}. 
DLC:
Simulated light curve of the satellite with
uniform surface but with dichotomy in reflectance properties on the leading
and trailing sides.
MLC: Simulated light
curve using albedo variations inferred by maps by the USGS and surface without dichotomy
(Section~\ref{subsec:Albedo}).
 }
    \label{fig:L4curve}
\end{figure}

\citet{DomingueVerbiscer1997} derived Hapke's rough surface parameters \citep{Hapke1984} 
from disc integrated intensities using Voyager and ground based data
for Europa, Ganymede and Callisto. The derived parameters indicate
differences in scattering characteristics of the leading and trailing hemispheres. This
dichotomy will contribute to the rotational light curve of the satellite.
Further, albedo variations due to distributed features are evident in the USGS mosaics;
these features will also determine the shape of the rotational light curve. 
In order to assess the relative contribution of surface dichotomy and albedo variation,
we simulated the rotational light curves of the Galilean satellites shown in Figure~\ref{fig:L4curve}, 
for a  solar phase angle of $6^\circ$ for the following two scenarios:
\begin{enumerate}
\item 
Uniform surface but for dichotomy in reflectance properties (DLC: short dashed line):
The light curves were constructed by summing the contribution from
$1^\circ \times 1^\circ$ elements on the visible surface
at orbital longitudes $0^\circ - 360^\circ$.
Hapke's parameters by \citet{DomingueVerbiscer1997} for the leading and trailing sides
in the 0.55 $\mu m$ wavelength band were used to estimate limb darkening on the satellite disc.
For Io as only global values 
of Hapke's parameters reported by \citet{McEwen1988} are available, this simulation could not be 
carried out.
\item 
Albedo variations inferred by Voyager I \& II and Galileo imagery (MLC: long dashed line): 
No dichotomy in scattering properties on the leading and trailing sides was considered.
The albedo maps constructed by various teams
were downloaded from the website of United States Geological Survey (USGS)\footnote{ 
Details of the photometric correction processes and mosaic construction are described
by \citet{Geissler1999} for Io, \citet{Phillips1997} for Europa, and \citet{Becker1999} 
for Ganymede and Callisto.}. For Io the {\it NIR} filter mosaic was used to suit our {\it R} 
band observations
while for the other satellites green filter mosaics were used. 
Hapke's parameters derived by \citet{McEwen1988} for orange filter were used to 
estimate the limb darkening 
for Io. For the other satellites average values of the Hapke's parameters for the leading
and trailing sides by \citet{DomingueVerbiscer1997} were used.  
The MLC light curves closely resemble the simulations by \citet{EmelyanovGilbert2006}
in their Figures (1-4), except for some deviation especially for Callisto on the trailing side.
This discrepancy may be attributed to the fact that the mosaics used in the two investigations
are not the same.
\end{enumerate}
The solid line OLC in Figure~\ref{fig:L4curve} is the observed V magnitude rotational light curve from 
\citet{MorrisonMorrison}.
The light curves in the magnitude scale are shifted to coincide at 0$^\circ$ Central
Meridian Longitude (CML) for
a better visual comparison, although a  critical assessment rests more
on the gradient of the light curves. 
The prime meridian of the Galilean satellites intersects the equator at the sub-Jovian point. The
satellites being in synchronous rotation, the CML follows the orbital longitude and the rotational angle.

A comparison of the groundbased light curve with the simulated light curves 
shows that:
\begin{enumerate}
\item For Io, OLC can be closely approximated by MLC.
\item For Europa OLC runs close to both MLC and DLC. 
\item There is significant deviation of OLC from
MLC as well as DLC for Ganymede especially at eastern and western elongations. 
\citet{Helfenstein1985} derived Hapke's constants for selected regions on Ganymede.
The simulated light curves will match closely with the observed rotational light curves if
Hapke's constants for more such regions are available and used to estimate limb darkening, 
instead of using just the albedo map. 
However as the CML of Ganymede during
the 1E3/1O3 and 2E3/2O3 events in 2014-15 were in the range 350$^\circ-0^\circ- 31^\circ$ and
DLC matches closely with OLC in this region, we selected DLC 
distribution for Ganymede in our fits to the mutual event data.  
\item The CML of Callisto when occulted/eclipsed during the present apparition is 
$345^\circ-360^\circ-21^\circ$. In this CML range DLC deviates considerably from OLC 
indicating significant contribution from distributed bright features. 
\citet{Moor1999} describe the surface of Callisto to the first order, to be either bright (frost with 
geometric albedo $\approx$ 0.8) or dark (geometric albedo $\approx$ 0.2). 
As seen in the USGS maps, there are prominent bright regions scattered East of 0$^\circ$ longitude on 
Callisto. 
\citet{DomingueVerbiscer1997} determined
Hapke's parameters from disc integrated intensities. 
It is therefore reasonable to expect that the effect of the bright regions will be to add to the ambient
intensity represented by 
 DLC in the 0$^\circ$-180$^\circ$ region. 
We therefore consider two basic cases for Callisto: (1) The DLC model and (2) Convolution of DLC with 
the mosaic from
USGS : DLC$^M$. The DLC$^M$ model utilizes the Hapke's parameters for the leading and trailing
sides to compute the limb darkening (like DLC) but in addition takes into account the albedo
map. The same set of Hapke's constants are used for the bright regions.
This is a simplistic approach. The center to limb variation in the intensity
of the bright frosty regions will not be the same as that of the average terrain due to
differences in surface compactness, grain sizes, local slope etc. In absence of
information on detailed scattering characteristics of the frosty regions,
we adopt the simple scenario mentioned above for DLC$^M$.
\end{enumerate}
 In our chosen model hereafter referred to as "Model 1" we adopt the distributions
MLC for Io, DLC for Europa, Ganymede and Callisto. For Callisto we use additionally
DLC$^M$ as discussed above.
\subsection{Extraction of the parameters}
\label{subsec:Extract}
The observed light curves were fitted with theoretical light curves computed using
the model presented by \citet{Vasundhara1994}. 
The fit involves comparing the observed (O) and computed (C) light curves  
using Marquardt's technique. 
We define the event plane to pass through the geometric center (GC) of the eclipsed/occulted
satellite S2. This plane is taken to be perpendicular to the extended heliocentric/topocentric vector to
GC of the eclipsed/occulted satellite intersecting  the event plane at S1. 
The two orthogonal parameters that were fitted are :
\begin{enumerate}
\item The correction to the impact parameter 
"$\Delta IP$", where $IP$ is the closest distance of the 
eclipsed/occulted satellite S2 from S1. 
\item The shift "$s$" of S2 along the track.
A positive value of $s$ indicates that the observed position of S2
is East of its predicted position. 
\end{enumerate}
 The observed time of close approach of S2 to S1 is obtained from
\begin{equation}
    T_{Obs}= T_{Pred} - s/v,
        \label{eq:Eq2}
\end{equation}
where $v$ is the projected relative velocity of S2 with respect to S1 on the event plane.
For an easterly motion of S2 with respect to S1 and positive $s$, the event will be in advance compared to
the prediction.
The observed $IP$ at the instant $T_{Obs}$ is given by
\begin{equation}
IP= IP_{Pred} +\Delta IP.
        \label{eq:IPobs}
\end{equation}
The separation (X,Y) of S2 from S1 at $T_{Obs}$
along EW and NS on the event plane are calculated from 
\begin{equation}
X=IP \times \sin P;~~
Y=IP \times \cos P, 
        \label{eq:XandY}
\end{equation}
where $P$ is the position angle of jovian pole of the date \citep{Archinaletal2011}.
We assume here that the apparent track of S2 relative to S1 is perpendicular to
the pole direction of the planet. 
The derived values of X and Y are independent of the theory. 
These can therefore be
used along with other kind of observations for updating the dynamical constants of the
satellites.
The residuals (O-C)X and (O-C)Y with respect to L2 ephemerides are given by:
\begin{equation}
    (O-C)X  =   s \times \cos P + \Delta IP \times \sin P 
        \label{eq:Eq5}
\end{equation}
\begin{equation}
    (O-C)Y  =  -s \times \sin P + \Delta IP \times \cos P.
        \label{eq:Eq6}
\end{equation}
Tables \ref{tab:O-table} and \ref{tab:E_table} give the results of the fits. 
The first two columns give the UT Date  and the event information. The third column gives
the intensity distribution that was used in the fit (Section \ref{subsec:Albedo}).
Column four gives the fitted observed time $T_{Obs}$ from Equation \ref{eq:Eq2}.
The derived value of X and Y at the instant $T_{Obs}$ from Equation~\ref{eq:XandY} are
given in columns five and six. Columns seven and eight give 
$(O-C)$ in X and Y from Equations~\ref{eq:Eq5} and \ref{eq:Eq6} respectively.
The central meridian longitude of the occulted/eclipsed 
satellite (CML$_2$) and the solar phase angles are given in the ninth and tenth columns
respectively. Column
eleven  gives the derived impact parameter in arcsec at $T_{Obs}$.
The impact parameter in km is given in column twelve.
The fitted light curves  and (O-C) residuals of the data points are given in the online version.
\begin{table*}
	\centering
	\caption{Results for Occultation events: Fitted observed time, relative inter-satellite distance and $(O-C)$ compared to L2 }  
	\label{tab:O-table}
	\begin{tabular}{cllrrrrrrrrr} 
		\hline
    UT Date      & Event&Surface&$T_{Obs}$(UTC)& $X$& $Y$&(O-C)X& (O-C)Y& $CML2$& Phase& \multicolumn{2}{c}{$IP$}\\
             &&model      &$\sigma$ $T_{Obs}$ &      &      &$\sigma$(O-C)X&$\sigma$(O-C)Y&&& \multicolumn{2}{c}{$\sigma(IP)$}\\
         &    &  &   hh mm ss.ss& arcsec&arcsec&arcsec &arcsec&Deg& Deg&arcsec&Km \\
(1)&(2)&(3)&(4)&(5)&(6)&(7)&(8)&(9)&(10)&(11)&(12)\\
		\hline
  2014 11 27 & 2O4&DLC & 22  8 43.81 &   0.0764 &   0.1978 &   0.0862 &   0.0907   &  18.7 &-10.4&   0.2120 &   770.2\\
             &     &&          1.23 &          &           &   0.0051 &   0.0076 &    &    &      0.0080&  28.9\\
  2014 11 27 & 2O4&DLC$^M$ & 22  8 57.88 &   0.0400 &   0.1034 &   0.0009 &   0.0153   &  18.7 &-10.4&   0.1109 &   402.9\\
             &     &&          1.23 &          &           &   0.0066 &   0.0132 &    &    &      0.0141&  51.2\\
  2015 01 31 & 2O1&MLC & 19 51 48.73 &  -0.1383 &  -0.3785 &  -0.0141 &   0.0048   & 282.8 & -1.3&  -0.4030 & -1271.9\\
             &     &&          0.34 &          &           &   0.0015 &   0.0012 &    &    &      0.0012&   3.7\\
  2015 02 25 & 2O1&MLC & 15 10  1.16 &  -0.0219 &  -0.0631 &  -0.0077 &  -0.0183   & 293.2 &  3.9&  -0.0668 &  -213.6\\
             &     &&          0.31 &          &           &   0.0025 &   0.0056 &    &    &      0.0059&  19.0\\
  2015 02 26 & 4O2&DLC & 20 27 57.79 &  -0.3066 &  -0.8837 &  -0.0348 &  -0.0177   & 340.7 &  4.1&  -0.9354 & -2996.1\\
             &     &&          0.76 &          &           &   0.0048 &   0.0023 &    &    &      0.0017&   5.6\\
  2015 03 11 & 2O1&MLC & 19 20 44.61 &   0.0654 &   0.1929 &  -0.0055 &   0.0041   & 298.6 &  6.5&   0.2037 &   667.4\\
             &     &&          0.28 &          &           &   0.0017 &   0.0020 &    &    &      0.0020&   6.7\\
  2015 04 04 & 1O3&DLC & 15  0 44.39 &   0.2020 &   0.6076 &  -0.0007 &  -0.0093   &  22.1 &  9.5&   0.6403 &  2234.5\\
             &     &&          0.81 &          &           &   0.0034 &   0.0023 &    &    &      0.0021&   7.3\\
  2015 04 11 & 1O3&DLC & 18  0 42.97 &   0.2158 &   0.6491 &   0.0077 &  -0.0063   &  20.6 & 10.1&   0.6841 &  2438.8\\
             &     &&          2.01 &          &           &   0.0097 &   0.0062 &    &    &      0.0056&  20.1\\
  2015 06 03 & 3O1&MLC & 15 31 31.50 &   0.0654 &   0.1829 &   0.0047 &   0.0071   &  74.5 &  9.8&   0.1942 &   807.0\\
             &     &&          0.75 &          &           &   0.0030 &   0.0046 &    &    &      0.0048&  19.8\\
		\hline
		\hline
	\end{tabular}
\end{table*}
\begin{table*}
        \centering
	\caption{Results for Eclipse  events: Fitted observed time, relative inter-satellite distance and $(O-C)$ compared to L2 }  
        \label{tab:E_table}
        \begin{tabular}{cllrrrrrrrrr} 
                \hline
    UT Date      & Event&Surface&$T_{Obs}$(UTC)& $X$& $Y$&(O-C)X& (O-C)Y& $CML2$& Phase& \multicolumn{2}{c}{$IP$}\\
             &      &model&$\sigma$ $T$ &\multicolumn{2}{c}{S2$-$S1} &$\sigma$(O-C)X&$\sigma$(O-C)Y&&& \multicolumn{2}{c}{$\sigma(IP)$}\\
   Date      &      &&   hh mm ss.ss& arcsec&arcsec&arcsec &arcsec&Deg& Deg&arcsec&Km \\
(1)&(2)&(3)&(4)&(5)&(6)&(7)&(8)&(9)&(10)&(11)&(12)\\
		\hline
  2015 01 26 & 3E1&MLC & 15 28 36.86 &   0.2802 &   0.7873 &  -0.0104 &   0.0107   & 309.6 & -2.3&   0.8357 &  3230.8\\
             &     &&          0.86 &          &           &   0.0045 &   0.0022 &    &    &      0.0017&   6.4\\
  2015 01 29 & 1E3&DLC & 16 13 36.33 &   0.2594 &   0.7259 &   0.0088 &  -0.0026   & 350.0 & -1.7&   0.7708 &  2983.6\\
             &     &&          0.79 &          &           &   0.0051 &   0.0028 &    &    &      0.0023&   9.0\\
  2015 01 31 & 2E1&MLC & 19 36 11.61 &   0.1097 &   0.3062 &  -0.0059 &   0.0039   & 281.8 & -1.3&   0.3252 &  1257.4\\
             &     &&          0.44 &          &           &   0.0016 &   0.0012 &    &    &      0.0012&   4.6\\
  2015 01 31 & 2E4&DLC & 21  4 13.84 &  -0.3581 &  -0.9998 &  -0.0044 &   0.0212   & 345.6 & -1.2&  -1.0620 & -4115.0\\
             &     &&          3.47 &          &           &   0.0150 &   0.0068 &    &    &      0.0045&  17.5\\
  2015 01 31 & 2E4&DLC$^M$ & 21  4 14.47 &  -0.3595 &  -1.0036 &  -0.0084 &   0.0184   & 345.6 & -1.2&  -1.0660 & -4130.7\\
             &     &&          3.34 &          &           &   0.0144 &   0.0068 &    &    &      0.0048&  18.5\\
  2015 02 12 & 1E3&DLC & 21 47 31.50 &   0.1353 &   0.3721 &   0.0154 &   0.0079   & 344.9 &  1.3&   0.3960 &  1533.8\\
             &     &&          0.51 &          &           &   0.0031 &   0.0027 &    &    &      0.0026&  10.3\\
  2015 02 18 & 2E1&MLC & 13 34 30.27 &   0.0515 &   0.1408 &   0.0374 &   0.0646   & 291.9 &  2.5&   0.1499 &   580.3\\
             &     &&          1.18 &          &           &   0.0057 &   0.0077 &    &    &      0.0080&  30.9\\
  2015 02 25 & 2E1&MLC & 15 55  4.28 &  -0.0107 &  -0.0290 &   0.0101 &  -0.0024   & 295.6 &  3.9&  -0.0309 &  -119.6\\
             &     &&          0.10 &          &           &   0.0010 &   0.0025 &    &    &      0.0026&  10.2\\
  2015 03 07 & 2E4&DLC & 14 58 32.26 &  -0.1590 &  -0.4257 &   0.0497 &  -0.0410   &  12.1 &  5.7&  -0.4544 & -1764.5\\
             &     &&          0.68 &          &           &   0.0033 &   0.0027 &    &    &      0.0026&  10.0\\
  2015 03 07 & 2E4&DLC$^M$ & 14 58 38.97 &  -0.1420 &  -0.3803 &   0.0358 &   0.0159   &  12.1 &  5.7&  -0.4060 & -1576.3\\
             &     &&          0.71 &          &           &   0.0034 &   0.0031 &    &    &      0.0030&  11.7\\
  2015 03 11 & 2E1&MLC & 20 32 59.37 &  -0.0952 &  -0.2538 &   0.0027 &  -0.0076   & 302.3 &  6.5&  -0.2711 & -1050.6\\
             &     &&          0.27 &          &           &   0.0014 &   0.0013 &    &    &      0.0012&   4.8\\
  2015 03 21 & 2E3&DLC & 15 45 59.29 &   0.2748 &   0.7246 &   0.0084 &  -0.0352   &  30.9 &  7.9&   0.7750 &  3007.8\\
             &     &&          0.87 &          &           &   0.0037 &   0.0019 &    &    &      0.0015&   5.7\\
  2015 03 24 & 2E4&DLC & 18 52 55.72 &  -0.2350 &  -0.6173 &   0.0534 &   0.0056   &  20.8 &  8.3&  -0.6605 & -2566.7\\
             &     &&          1.36 &          &           &   0.0020 &   0.0011 &    &    &      0.0009&   3.6\\
  2015 03 24 & 2E4&DLC$^M$ & 18 53 17.48 &  -0.2317 &  -0.6088 &   0.0251 &   0.0260   &  20.8 &  8.3&  -0.6514 & -2531.5\\
             &     &&          1.30 &          &           &   0.0019 &   0.0012 &    &    &      0.0010&   3.7\\
  2015 03 28 & 1E3&DLC & 15  3 36.79 &  -0.2122 &  -0.5555 &   0.0010 &   0.0162   &  21.1 &  8.8&  -0.5947 & -2309.2\\
             &     &&          0.29 &          &           &   0.0012 &   0.0008 &    &    &      0.0007&   2.7\\
  2015 03 28 & 2E3&DLC & 19  6 54.20 &   0.2045 &   0.5351 &   0.0268 &  -0.0352   &  29.5 &  8.8&   0.5729 &  2224.5\\
             &     &&          0.78 &          &           &   0.0035 &   0.0023 &    &    &      0.0020&   7.8\\
  2015 04 02 & 4E3&DLC & 19 13  2.18 &  -0.4442 &  -1.1580 &  -0.0145 &   0.0372   & 281.3 &  9.3&  -1.2403 & -4812.7\\
             &     &&         12.67 &          &           &   0.0274 &   0.0125 &    &    &      0.0073&  28.4\\
  2015 04 05 & 2E1&MLC & 16 32 23.36 &  -0.2583 &  -0.6709 &  -0.0093 &   0.0008   & 312.9 &  9.6&  -0.7189 & -2790.2\\
             &     &&          1.45 &          &           &   0.0084 &   0.0039 &    &    &      0.0025&   9.5\\
		\hline
	\end{tabular}
\end{table*}

The standard deviations of the derived parameters reported in
Table~\ref{tab:O-table} and Table~\ref{tab:E_table} are derived from 
$\sigma \Delta IP$ and $\sigma s$ calculated using the covariance 
matrix.
These estimates therefore depend on the noise which has Poisson distribution, in the data. 
 Other factors such as error in the back ground subtraction
due to proximity of Jupiter or fluctuating sky transparency may
produce much larger uncertainty in the derived parameters. 
The R.M.S. residuals of the mean value
of the estimated parameters are determined from the entire data set as shown in the next section.
Another source of uncertainty arises because
our observations are in R band, due to unavailability of albedo map for Callisto in red
filter we used the green filter albedo map. 
Similarly, use of Hapke's parameters for Europa, Ganymede and Callisto in the closest available
band at 0.55$\mu m$ is cause for some concern.
\section{Results and discussions}
\label{sec:Results}
Results of the fits using Model 1 for the eight occultation events
and fifteen eclipse events are given
in Tables \ref{tab:O-table} 
and \ref{tab:E_table} respectively. 
The time of closest approach of the satellites and the relative astrometric positions
at this instant are the actual observed positions, topocentric for occultations and heliocentric
for eclipses. These are independent of the theory and can be used along with other sets of observations
in construction of future ephemerides.

We consider the two orthogonal parameters 
$\Delta IP$ and
$s$ for comparing the observations
with predictions instead of the projections (O-C)X and (O-C)Y computed using
Equations ~\ref{eq:Eq5} and \ref{eq:Eq6}. 
 
\begin{figure}
        \includegraphics[width=3.5in]{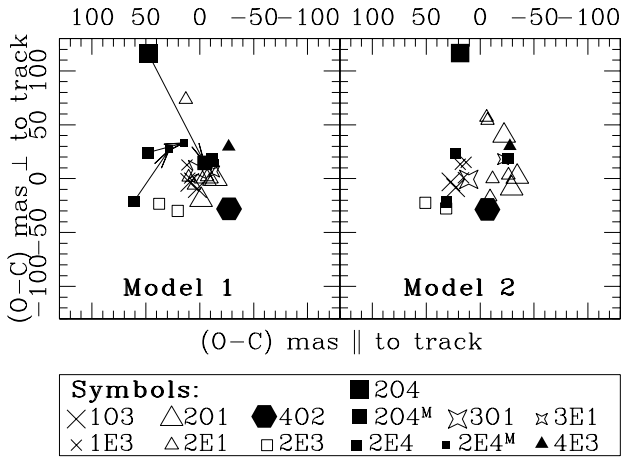}
\vskip -2.5cm
    \caption{The derived (O-C) residuals along and perpendicular
to the track for Models 1 and 2. Model 1 uses the intensity distributions
MLC for Io and DLC for Europa, Ganymede and Callisto. For Callisto,  DLC$^M$ refers to
albedo map + dichotomy in leading and trailing sides. Model 2 uses a uniform disc
model without albedo variation except for limb darkening. 
Symbols for the event pairs are given in the box. 2O4$^M$ and 2E4$^M$ are fitted positions
using DLC$^M$ distribution.}
    \label{fig:L1L2_figure}
\end{figure}

\begin{figure*}
        \includegraphics[width=8.6in]{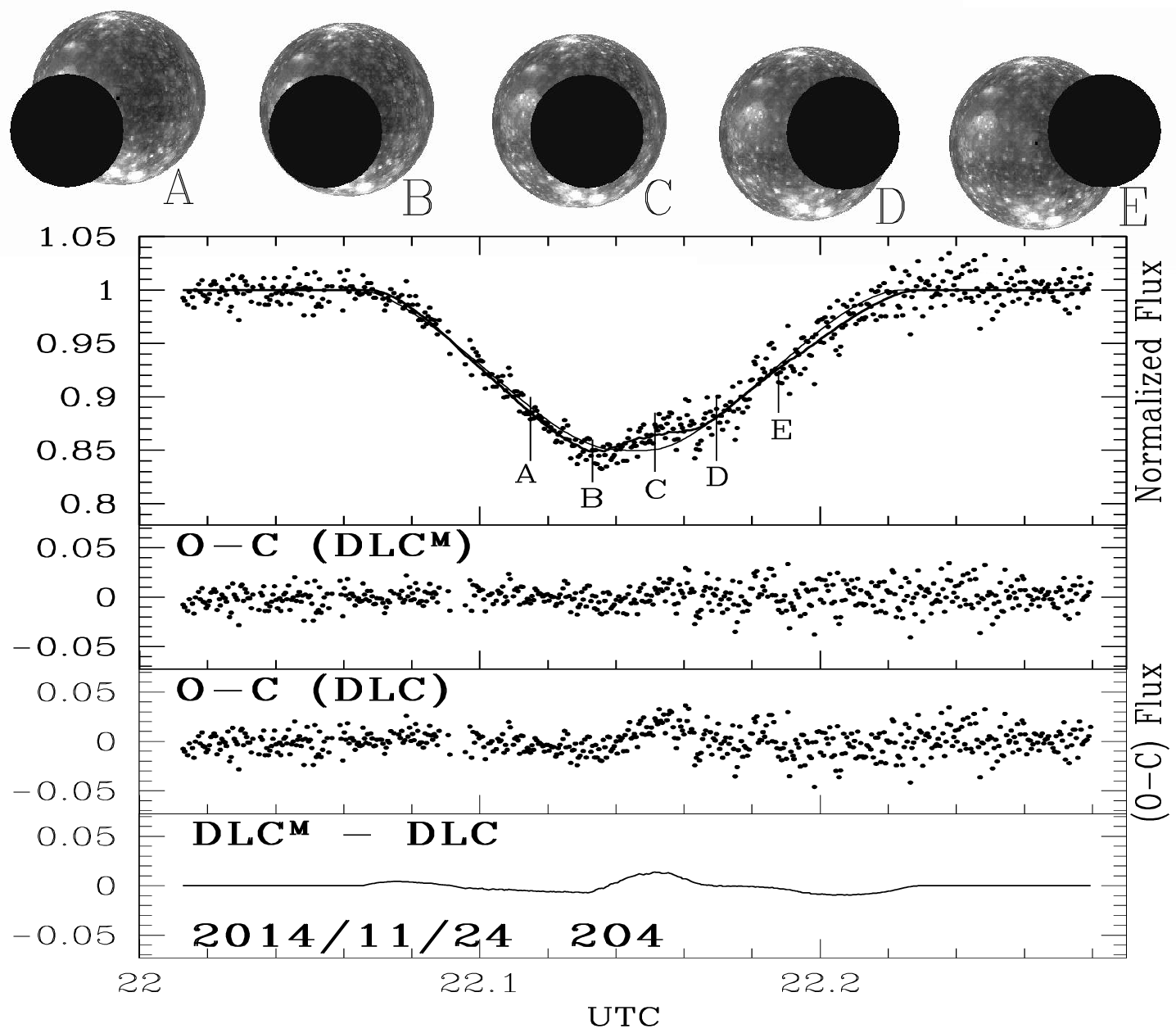}
\vskip -16cm
    \caption{Top panel in the box: Fitted light curves using DLC(Thin solid line) 
and DLC$^M$(Thick solid line) of the 2O4 event on 27 November 2014. 
The relative positions of the two satellites
at instances 'A', 'B','C', 'D' and 'E' marked on the light curve are shown above the
box.  The eastern and South-East
segment on the disc of Callisto dominated by the bright regions Valhalla, Adlinda,
Lofn and Heimdall is responsible for asymmetry of the observed light curve.
The (O-C) residuals for DLC and DLC$^M$ are shown in the two middle panels. The lowermost panel shows
the contribution of the segment on the satellite with bright regions in explaining the
asymmetry. The O-C is significantly improved in case of DLC$^M$.}
    \label{fig:27N_figure}
\end{figure*}

To check the validity of our selection of
the limb darkening and albedo model (Model 1) in the present investigation, we 
fitted the light curves 
using another scenario (Model 2) in which:
\begin{enumerate}
\item The occulted/eclipsed satellite is  uniform, i.e. without albedo variation and also 
without surface dichotomy. 
\item  The satellite is limb darkened using average value
of Hapke's parameters of the leading and trailing sides. For Io the global average values
were used.
\end{enumerate}
The derived (O-C) residuals of $"s"$ (along the track) and  $"\Delta IP"$ (perpendicular
to the track) are shown in Figure~\ref{fig:L1L2_figure} for Models 1 and 2. 
Table \ref{tab:OMC-table} gives the mean and R.M.S. of the residuals  for
events of the same type that were observed  at least four times.
A large number of events are needed for an
unequivocal conclusion. However, some trends are evident in 
Figure \ref{fig:L1L2_figure} and Table~\ref{tab:OMC-table}: 
\begin{enumerate} 
\item For the 2E1/2O1 events the scatter in (O-C) perpendicular to track in case of 
Model 2 is significantly
larger compared to Model 1, except for the 2E1 event on 18 February observed
at large air mass (Table~\ref{tab:Obs_table}). This is also evident from column 5 of Table~\ref{tab:OMC-table}; 
the R.M.S. values are  7.8 and 19.3  for Models 1 and 2
respectively. Model 1 appears to constrain the estimation of impact parameter better than Model 2. 
However for the five 1E3/1O3 events which are partial events 
there are no significant differences either in the mean value or the R.M.S. of the residuals with the two
models. 
\item The mean and R.M.S. residuals along the track for the 2E1/2O1 events with Model 2 are $-19.9$ and $9.3$ mas
respectively implying that the fitted times are delayed compared to L2. The delay is marginal for fits  
with Model 1 which yields mean and R.M.S. residuals of $-3.3$ and $8.7$ mas respectively.
\citet{Vasundhara2002} showed that as a result of albedo variation, the light center on
Io can shift by $\approx$ 50 km ($\approx 14$ mas) towards West when near western elongation. 
As these events occurred when CMLs were in the  range $281^\circ - 313^\circ$, 
albedo variation may be a likely cause for the delay. Model 2
assumes the disc to be uniform; the deepest part of the light curve is interpreted as the close
approach instant of the GC of S2 to S1 thus delaying the fitted time.
Model 1 intrinsically takes into account the albedo variation and hence is reliable.
\item The mean (O-C) residuals along the track in case of the five 1E3/1O3 events for fits
with Models 1 and 2 are 5.6 and 19.3 mas respectively. Model 2 predicts the event to be in advance
compared to Model 1;  Model 2
detects Ganymede to be east of its actual position. The events
occurred within $\pm 25^\circ$ of the superior conjunction of Ganymede. As the leading side is brighter
than the trailing side for Ganymede, the light center of its eclipsed/occulted portion will be
to the East of the GC of the disc. 
Model 1 which takes into account this dichotomy has a lower residual while Model 2
interprets the light center as the GC and hence the event fitted time is in advance.  
\item The (O-C) residuals along the track of the two 2E3 events in Figure~\ref{fig:L1L2_figure} 
have large mean residuals of  29 mas
and 41 mas for Models 1 and 2 respectively.
The data sample being sparse no decisive conclusion can be drawn.
\item As shown by the arrows connecting DLC to DLC$^M$ positions, the (O-C) residuals of the 2E4/2O4 
events improve 
when the albedo map of Callisto is included. 
Figure~\ref{fig:27N_figure} shows the  fitted light curves and (O-C) using DLC and DLC$^M$
distributions for the 2O4 event on 27 November 2014. 
The observed light curve is asymmetric. Distribution DLC (thin solid line) 
fits only approximately the central 
portion of the light curve, the (O-C) plot in this region indicates  poor fit with a hump indicating
unaccounted flux.
DLC$^M$ (thick solid line) fits better with smaller (O-C). The relative position of the two satellites
at instances 'A', 'B','C', 'D' and 'E' marked on the light curve are shown at the top.
From the image of Callisto in 'E' it is clear that the eastern
segment is dominated by the bright regions: Hepti, impact basin Valhalla, Adlinda, 
 Lofn and Heimdall. Starting from the first contact, the south-eastern bright regions start getting occulted. 
Between positions 'A' and second contact (position B) the southern part of the segment with bright 
regions except the southernmost feature Heimdall are occulted. This explains the relatively rapid fall in 
intensity in this part of the 
light curve compared to DLC. While DLC distribution yields a light curve that
is flat between second and third contacts except for limb darkening effects, DLM$^M$ produces
intensity variations in this region. The rise between 'B' and 'C' is due to unveiling
of the southern bright group. This is demonstrated in the lower most panel DLC$^M$ - DLC.  
For annular events the use of albedo map thus appears to improve 
the fits.
\end{enumerate}

\begin{table}
        \centering
	\caption{Comparison of(O-C) residuals with the two models}  

        \label{tab:OMC-table}
        \begin{tabular}{lrrrrcc} 
                \hline
 Ephm. & \multicolumn{2}{c}{$\|$ Track}     & \multicolumn{2}{c}{$\bot$ to Track} &Event&N \\
       & \multicolumn{2}{c}{$(O-C)$}     & \multicolumn{2}{c}{$(O-C)$}  \\
       & $Mean$ &R.M.S.& Mean &R.M.S.&J1~J2\\
                \hline
(1)&(2)&(3)&(4)&(5)&(6)&(7)\\
                \hline
Model 1&     -3.3&     8.7&    -3.4&     7.8  &2~~1   &7\\
Model 2&     -19.9&     9.3&    10.4&    19.3  &   &7\\
                \hline
Model 1 &     5.6&     6.8&     3.2&     10.5  &1~~3   &5\\
Model 2 &     19.3&     5.7&     3.6&    10.3   &    &5\\
                \hline
Model 1$^1$  &     5.9&     18.9&     22.4&     9.5  &2~~4   &4\\
Model 2 &     11.5&    25.4&    34.3&    34.0  &   &4\\
                \hline
                \hline
\multicolumn{7}{l}{1. With DLC$^M$ albedo distribution}
	\end{tabular}
\end{table}

\section{Concluding remarks}
We use our observations of mutual events of the Galilean satellites in 2014-15
from VBO to show that a robust fit to the light curves 
depends on precise knowledge of intensity distribution on eclipsed/occulted satellites.  We set the criterion that a satisfactory distribution
must also explain the observed rotational light curve of the satellites.
For Io, albedo map by USGS along with global
limb darkening constants by \citet{McEwen1988} is found to be satisfactory.
Dichotomy in the scattering characteristics 
of the leading and trailing sides \citep{DomingueVerbiscer1997} 
is able to approximate the observed rotational light curves for the three outer satellites.
The departures which
are more severe for Ganymede and Callisto can
be reduced if Hapke's parameters for prominent terrains on the satellites
are available. These parameters will help to estimate the  center to limb variation of intensity of the 
local terrains at different rotational phases. 

Our results with uniform disc model (Model 2) indicate  bias in the (O-C) residuals along the track for sets
of 2E1/2O1 and 1E3/1O3 events in the direction and extent that is consistent with the shift
of light center on Io and Ganymede respectively. No such bias is evident within the R.M.S. limits
for results using our chosen intensity distribution model (Model 1).
As the events of a given kind occur within a narrow range of orbital longitudes,
such a bias will be wrongly interpreted as correction in longitude of either of the satellites.
Right choice of a model is therefore important because the theory 
is already able to make predictions accurate to a few mas \citep{Lainey2009} and observational
techniques have improved enormously.

The time of closest approach of the satellites and the relative astrometric positions
at this instant are presented for twenty three events. 
\section*{Acknowledgments}
We thank the J. C. Bhattacharyya telescope time allocation Committee 
for liberally allocating time for these observations.

\bsp	
\label{lastpage}
\end{document}